\documentclass[graybox]{svmult}

\usepackage{mathptmx}       % selects Times Roman as basic font
\usepackage{helvet}         % selects Helvetica as sans-serif font
\usepackage{courier}        % selects Courier as typewriter font
\usepackage{makeidx}         % allows index generation
\usepackage{graphicx}        % standard LaTeX graphics tool
\usepackage{multicol}        % used for the two-column index
\makeindex             % used for the subject index
\begin{document}

\title*{Period-luminosity relationship for SX Phoenicis stars in Galactic globular clusters}
\author{Grzegorz Kopacki and Andrzej Pigulski}
\institute{Grzegorz Kopacki and Andrzej Pigulski 
 \at Instytut Astronomiczny Uniwersytetu Wroc\l{}awskiego, Kopernika 11, 51-622 Wroc\l{}aw, Poland,
 \email{kopacki@astro.uni.wroc.pl}
}
\maketitle
% Please use both starred abstract and non-starred abstract.

\abstract*{We compiled a list of about 250 SX Phoenicis stars known in Galactic globular 
clusters in order to study period-luminosity relation for this type of 
variable. The absolute magnitudes of these stars are derived using metallicity-luminosity 
calibration for RR Lyrae stars. The mixture of different radial and non-radial 
modes present in SX Phoenicis stars and the lack of unique method of mode 
identification cause the difficulties in defining strict period-luminosity 
relation. As a solution we propose to use confirmed 
double-mode radial pulsators.}

\abstract{We compiled a list of about 250 SX Phoenicis stars known in Galactic globular 
clusters in order to study period-luminosity relation for this type of 
variable. The absolute magnitudes of these stars are derived using metallicity-luminosity 
calibration for RR Lyrae stars. The mixture of different radial and non-radial 
modes present in SX Phoenicis stars and the lack of unique method of mode 
identification cause the difficulties in defining strict period-luminosity 
relation. As a solution we propose to use confirmed 
double-mode radial pulsators.}

\section{Introduction, Database and Results}

Although intrinsically fainter than Cepheids or RR Lyrae stars, SX Phoenicis
variables are gaining increasing attention, because of their possible usage as distance
indicators through the period -- luminosity (PL) relation. They are short-period pulsating
stars considered as a Population II counterparts of $\delta$ Scuti stars and are found
frequently in globular clusters (GCs). The existence of the PL relation for SX Phoenicis stars
has been clearly established both theoretically \cite{San01} and 
empirically (e.g.\ \cite{Por08}). 

We performed an extensive search for SX Phoenicis stars in Galactic GCs 
searching the literature, a little-bit obsolate 
catalogue of SX Phoenicis stars in GCs \cite{Rod00} and revised but not complete
version of the Catalogue of Variable Stars in GCs \cite{Cle01}. Altogether, 
245 stars of this type with suitable photometric data located in 30 GCs were found including
19 double-mode radial pulsators. 

The absolute magnitudes of the studied stars were derived using metallicity-luminosity 
calibration for RR Lyrae stars \cite{Gra03}. The average
visual magnitudes of the horizontal branch stars and cluster metallicities were taken from
the Catalogue of Parameters for Milky Way GCs \cite{Har96}.

\begin{figure}[t]
\sidecaption[t]
\includegraphics{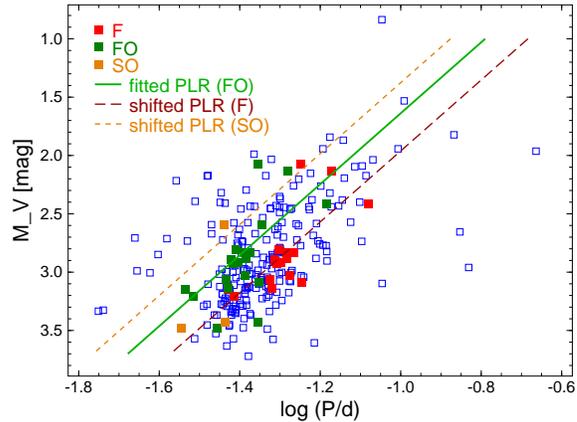}
\caption{The period -- luminosity diagram for analyzed SX Phoenicis stars.
 Double radial-mode stars are indicated with filled symbols. 
 The solid line shows PL relation (PLR) for the first overtone (FO) mode obtained
 from the linear fit for the observed F/FO and FO/SO stars excluding two faintest stars
 (FO/SO pulsators from $\omega$ Cen), which seem to be too faint for their periods.}
\label{fig:1}       % Give a unique label
\end{figure}

The resulting PL relation for all studied SX Phoenicis stars is shown in Fig.\ \ref{fig:1}.
The scatter is very large and no strict relation(s) can be defined. 
The main reason for this is surely the presence of stars pulsating in different modes, both
radial and non-radial. 
%Moreover, the large photometric amplitude is a poor discriminant of 
%the radial pulsation mode in SX Phoenicis stars. 

The PL relation can be defined only for stars pulsating in the same radial mode. Consequently, we first
need to identify modes observed in SX Phoenicis stars. Unfortunately,
there is no unique method of mode identification based on simple photometric
parameters derived solely from the light curve. As a solution to this problem we made use of
confirmed double-mode radial pulsators which can be quite easily identified through their period
ratios.  Now, evident separation between
fundamental (F) and first overtone (FO) modes defining two parallel strips can be discerned in Fig.\ \ref{fig:1}.
Fitting line to the FO components
yields the slope of $-3.04$ which is very 
close to the recent determinations of this parameter (e.g.\ \cite{Coh11}).

%(standard deviation of the fit $\sigma=0.2$ mag)

\end{document}